\documentstyle[twoside,fleqn,espcrc2,epsf]{article}

\newcommand{\AmS}{{\protect\the\textfont2
  A\kern-.1667em\lower.5ex\hbox{M}\kern-.125emS}}

\title{Top quark production near threshold\thanks{%
       Talk presented at the High Energy Physics 
       International Euroconference on Quantum Chromodynamics (QCD'99), 
       Montpellier, France, 7-13 July 1999.}
}

\author{
\vspace{-3.5cm}
\begin{flushright}
  CERN-TH/99-281\\
  September 1999\\
  hep-ph/9910534
\end{flushright}\vspace{2.0cm}
M. Beneke\address{Theory Division, CERN, CH-1211 Geneva 23, 
        Switzerland}}

\newcommand{\bff}[1]{\mbox{\boldmath ${#1}$}}

\begin{document}

\begin{abstract}
\noindent
The present theoretical status of top quark pair production near 
threshold at (future) $e^+ e^-$ ($\mu^+\mu^-$) colliders is summarized.
\end{abstract}

\maketitle

\thispagestyle{empty}

\section{INTRODUCTION}

Only a few of the top quark's properties are currently known. In 
the future the Tevatron and the LHC, and a Linear Collider, will measure 
many of its couplings in great detail. This will tell us whether the 
top quark is just another standard model quark (in which case it would 
be a rather uninteresting one) or whether it plays a special role  
in physics beyond the standard model, as perhaps its large mass suggests.

The measurement of the top quark mass and decay width through a scan of the 
top cross section near the pair production threshold is unique to a 
high-energy lepton collider -- and only indirectly related to 
non-standard model physics. (At a hadron collider the top quark 
pair invariant mass distribution is analogous to the threshold scan. 
However, the invariant mass can probably not be measured with an 
accuracy below $1\,$GeV as is necessary to make the subsequent discussion 
relevant.) Besides providing an accurate 
measurement of these parameters, the threshold 
scan probes a new and interesting regime of strong interaction dynamics: 
for a very short time between its birth and death through single top 
quark decay, the $t\bar{t}$ pair is the only strongly interacting 
system we know which is bound by a perturbatively calculable heavy 
quark potential. Contrary to charmonium and bottomonium, which are 
partly non-perturbative, toponium properties can be computed systematically.

Although leading order and next-to-leading order calculations of the 
threshold production cross section have been completed quite some time 
ago \cite{lo}, the methods to carry out such calculations systematically  
have been fully developed only recently. They have been used so far to 
compute the $t\bar{t}$ production cross section to next-to-next-to-leading 
order (NNLO) in the threshold region \cite{nnlo1,BSS99,nnlo2}. 
In this talk, based on \cite{BSS99}, I give a brief account of the 
calculational tools involved and discuss the surprising conclusions 
that emerged from the NNLO calculation.

\section{SYSTEMATICS}

The threshold region is defined by the scaling rule $v\sim \alpha_s(m_t v)$, 
where $v$ is defined through $E\equiv m_t v^2\equiv \sqrt{s}-2 m_t$ and 
$\sqrt{s}=q^2$ is the centre-of-mass energy squared. 
$m_t$ denotes the pole mass --  
mass renormalization is an important issue, which will be discussed in 
detail later. Since $v\ll 1$, several 
scales are relevant to $t\bar{t}$ production near the threshold. The 
approach described below treats these scales sequentially, working 
downwards from the largest scale $m_t$. To do this, one has to know what 
the relevant momentum regions are. Just as one distinguishes hard, collinear 
and soft particles in high-energy collisions of (massless) quarks and 
gluons, one can identify the following momentum regions relevant to 
non-relativistic heavy quarks \cite{BS98}:
hard (energy $l_0\sim m_t$, momentum $\bff{l}\sim m_t$ -- 
referring to a frame where $\bff{q}=0$) heavy quarks and gluons 
(light quarks), soft ($l_0\sim m_t v$, $\bff{l}\sim m_t v$) heavy 
quarks and gluons, 
potential ($l_0\sim m_t v^2$, $\bff{l}\sim m_t v$) heavy quarks and gluons 
and ultrasoft ($l_0\sim m_t v^2$, $\bff{l}\sim m_t v^2$) gluons. For top 
quarks the soft scale turns out to be of order $20\,$GeV, the ultrasoft 
scale of order $2\,$GeV -- large enough for a perturbative 
treatment. However, the presence of the small kinematic parameter $v$ 
implies that some terms have to be summed to all orders in $\alpha_s$.

\subsection{Coulomb resummation}

The exchange of a Coulomb (potential) gluon results in a `correction' of 
order $\alpha_s/v\sim 1$. The leading order Coulomb interaction must 
be treated non-perturbatively. Defining the $R$-ratio for the top production 
cross section through a virtual photon in the usual way, this leads to 
a succession of LO, NLO, ... approximations defined by keeping all 
terms of the form 
\begin{eqnarray}
\label{syst}
R&\equiv&\sigma_{t\bar{t}}/\sigma_{\mu^+\mu^-} = 
v \,\sum_k \left(\frac{\alpha_s}{v}\right)^k \cdot 
\big\{1\,\mbox{(LO)};
\nonumber\\
&&\hspace*{-0.5cm}
\alpha_s,v\,\mbox{(NLO)};\,
\alpha_s^2,\alpha_s v,v^2\,\mbox{(NNLO)};\,\ldots\big\}.
\end{eqnarray}
Note that at NNLO we do not need the exact perturbative coefficient 
of the $\alpha_s$-expansion, but only the first three terms in an expansion 
in $v$. This makes it useful to construct an expansion method 
(the `threshold expansion') which allows 
us to compute the expansion coefficients 
without knowing the exact result \cite{BS98}.

\subsection{Logarithms}

The resummation scheme (\ref{syst}) replaces the conventional 
$\alpha_s$-expansion. Each further order in the resummation improves 
the theoretical accuracy by one power of $\alpha_s$ as usual. But 
because of the very different momentum scales involved, large logarithms 
of $v$ and $v^2\sim 1/100$ are left over. These logarithms can also 
be summed using renormalization group methods. The leading logarithmic 
(LL), next-to-leading logarithmic (NLL), ... approximation is defined 
by keeping all terms of the form 
\begin{eqnarray}
\label{systrge}
R &=& 
v \,\sum_k \left(\frac{\alpha_s}{v}\right)^k 
\sum_l \left(\alpha_s \ln v\right)^l \cdot 
\big\{1\,\mbox{(LL)};
\nonumber\\
&&\hspace*{-0.5cm}
\alpha_s,v\,\mbox{(NLL)};\,
\alpha_s^2,\alpha_s v,v^2\,\mbox{(NNLL)};\,\ldots\big\}.
\end{eqnarray}
In practice, the summation of logarithms is done first by summing them 
into the coefficient functions of operators in the effective field theories  
discussed below. The summation of Coulomb $\alpha_s/v$ corrections 
is performed at the end by computing scattering amplitudes in the effective 
theory perturbatively.

\subsection{Treatment of the top quark width}

The decay $t\to Wb$ proceeds rapidly and the single top decay width 
cannot be neglected near threshold. In fact, in the standard model, 
$\Gamma_t\sim m_t v^2$. For hard and soft top quarks the width is a 
small correction to the quark propagator and can be expanded. For 
potential top quarks with energy $E\sim m_t v^2$ and momentum $\bff{p}\sim 
m_t v$, we can approximate the quark propagator 
\begin{eqnarray}
\frac{1}{\not\!P_t-m_t-\Sigma(P_t)}&\approx& 
\nonumber\\
&&\hspace*{-2.5cm}\frac{1}{E+i\Gamma_t-
\bff{p}^2/(2 m_t)} \,\left[1 + {\cal O}(v)\right].
\end{eqnarray}
The width is a leading order effect (which cannot be expanded), but 
can be taken into account at leading order by making the top quark 
energy complex: $E\to E+i\Gamma_t$, where $\Gamma_t$ is the 
gauge-independent on-shell decay width. Beyond leading order, 
counting $\Gamma_t\sim m_t v^2$, the self-energy has to be matched to 
better accuracy. The correction terms relate to the off-shell self-energy 
and carry electro-weak gauge-dependence. A complete NNLO result in 
the presence of a width that scales as above therefore includes 
electroweak vertex corrections as well as single resonant backgrounds 
and non-factorizable corrections to the physical $WWb\bar{b}$ 
final state. A systematic treatment of these complications has not been 
attempted yet.

\subsection{Non-perturbative effects}

On top of the resummed perturbative expansion there exist non-perturbative 
effects suppressed by powers of the QCD scale $\Lambda$. The leading 
non-perturbative correction to the $t\bar{t}$ cross section far from 
threshold is supposed to be described by the gluon condensate and scales 
as $(\Lambda/m_t)^4$. As $m_t v^2\gg \Lambda$, the operator product 
expansion remains valid near threshold, but the relevant scale is now 
$m_t v^2$ rather than $m_t$. Accounting for the velocity suppression 
of the top coupling to a dynamical gluon, we obtain the estimate 
$\delta R_{\rm NP}/R\sim v^2(\Lambda/(m_t v^2))^4$, which is very small 
indeed. It therefore seems justified to neglect non-perturbative effects, 
although a more detailed investigation would certainly be worthwhile.

Note that there is no room for phenomenological, non-perturbative 
modifications of the heavy quark potential, such as adding a linear 
term of order $\Lambda^2 r$, in the present approach. Since 
$m_t v^2\gg \Lambda$ there is no reason to assume that a non-perturbative 
gluon with momentum of order $\Lambda$ would give rise to an 
instantaneous interaction. Rather it would modify the propagation of 
ultrasoft gluons in the effective theory discussed below, while 
leaving the potential unmodified. Estimating the size of non-perturbative 
effects by adding $\Lambda^2 r$ to the potential would 
result in the over-estimate $\delta R_{\rm NP}/R\sim (\Lambda/(m_t v))^2$.

\subsection{Present status}

NNLO calculations have now been done by several groups 
\cite{nnlo1,BSS99,nnlo2}. It is known that leading logarithmic contributions 
are absent \cite{BSS99}. Some NLL and 
NNLL corrections are known, but a systematic implementation of the 
renormalization group is still missing. (As correctly pointed out in 
\cite{LMR99}, the summation of NLL effects in \cite{BSS99} is not complete, 
contrary to the statement made there.) There is a calculation of 
some potentially 
important NNNLO terms \cite{KP99}. Finite width effects have been 
treated only to {\em leading} order, i.e. using $E\to E+i\Gamma_t$ 
(as in \cite{lo}), or variants that have equivalent 
parametric accuracy. Some non-factorizable contributions have been 
studied near threshold \cite{PS97}. As already mentioned, 
non-perturbative corrections have not yet been estimated, but 
they are expected to be small.

In order to optimize an accurate top quark mass determination it 
is important to choose an appropriate mass renormalization scheme,  
different from the on-shell scheme \cite{Ben98} that has been used by 
convention until recently. The latest calculations \cite{BSS99,nnlo2} 
make use of such schemes, though different ones.

\section{CALCULATION IN THREE STEPS}

We sketch how the resummed threshold cross section is obtained. This 
is done by constructing, in two steps, an effective theory in which all 
modes except potential quarks and ultrasoft gluons are integrated out. 
The last step implies computing the cross section perturbatively 
in the effective theory. Although the cross section is ultraviolet and 
infrared finite, intermediate steps in this construction lead to 
infrared and ultraviolet divergences, which have to be regularized 
consistently. It is convenient and simple to use dimensional 
regularization. 

\subsection{Step 1: Matching on NRQCD}

We begin with integrating out hard (relativistic) quarks and gluons. 
Their contribution goes into the coefficient functions of the 
non-relativistic QCD (NRQCD) Lagrangian \cite{nrqcd} and into the 
coefficient functions of non-relativistic external currents. The effective 
Lagrangian is
\begin{eqnarray}
\label{nrqcd}
{\cal L}_{\rm NRQCD} &=& 
\psi^\dagger \left(i D^0+\frac{\bff{D}^2}{2 m_t}+i\Gamma_t
\right)\psi
\nonumber\\
&&\hspace*{-1.5cm}
 + \,\mbox{NNLO} + \mbox{antiquark terms } + \,{\cal L}_{\rm light},
\end{eqnarray}
where $\psi$ is the non-relativistic top quark field. Only the terms 
that contribute at leading order are written out explicitly. (The 
remaining ones can be found in \cite{BSS99}.) The bilinear term 
proportional to $\Gamma_t$ arises from matching the heavy quark self-energy 
to leading order and effects the substitution $E\to E+i\Gamma_t$ as 
discussed above. The leading order terms are not renormalized. Hence 
logarithms can appear only in the NNLO terms. 

The top quark vector coupling to the virtual photon in the effective theory 
is given by
\begin{equation}
\label{current}
\bar{t}\gamma^i t = c_1\,\psi^\dagger \sigma^i\chi - 
\frac{c_2}{6 m_t^2}\,\psi^\dagger \sigma^i (i\bff{D})^2\chi + 
\ldots,
\end{equation}
where the ellipsis refers to terms beyond NNLO and $\chi$ denotes the 
top anti-quark field. At NNLO, we can 
use $c_2=1$, while $c_1$ is needed at order $\alpha_s^2$. The two-loop 
contribution to $c_1$ has been computed in \cite{BSS98}. To 
leading-logarithmic, NLL, NNLL, ... accuracy one also needs the 
1-loop, 2-loop, 3-loop, ... anomalous dimension of the operator 
$\psi^\dagger \sigma^i\chi$. The 1-loop anomalous dimension vanishes, 
so there are no LL effects. The 2-loop anomalous dimension is 
known \cite{BSS98} and contributes NLL terms. In addition the NNLO couplings 
in (\ref{nrqcd}) mix into $\psi^\dagger\sigma^i\chi$ through ultraviolet 
divergent potential loops. Hence the LL renormalization of these 
couplings contributes NLL terms to the $t\bar{t}$ cross section.

\subsection{Step 2: Matching on PNRQCD}

The loop integrals constructed with the non-relativistic Lagrangian still 
contain soft, potential and ultrasoft 
modes. Near threshold, where energies are of order $m_t v^2$, only 
potential top quarks and ultrasoft gluons (light quarks) can appear as 
external lines of a physical scattering amplitude. Hence, we integrate out 
soft gluons and quarks and potential gluons (light quarks) and 
construct the effective Lagrangian for the potential top quarks and 
ultrasoft gluons (light quarks). Because the modes that are integrated out 
have large energy but not large momentum compared to the modes we keep, 
the resulting Lagrangian contains instantaneous, but spatially non-local 
interactions in addition to the local interactions inherited from 
the NRQCD Lagrangian. In the simplest case, the instantaneous 
interactions reduce to what is commonly 
called the `heavy quark potential'. This theory is appropriately termed  
potential non-relativistic QCD (PNRQCD) \cite{PS98}. Although 
both are integrated out together, it is necessary to distinguish 
potential gluons (which are off mass shell) and soft gluons (which 
are on-shell) to obtain an homogeneous (single scale) expansion 
near threshold \cite{BS98}. However, both contribute to the heavy 
quark potential and only their sum is gauge-invariant (see also 
the discussion in \cite{Ben98a}).

There is no further modification of the external vector current 
induced by matching on PNRQCD up to NNLO. For the effective 
Lagrangian we quote again only the leading order Lagrangian 
explicitly:
\begin{eqnarray}
\label{pnrqcd}
{\cal L}_{\rm PNRQCD}
 &=& 
\psi^\dagger \left(i \partial^0+\frac{\bff{\partial}^2}{2 m_t}+i\Gamma_t
\right)\psi 
\nonumber\\
&&\hspace*{-2.0cm}
+ \,\chi^\dagger \left(i \partial^0-\frac{\bff{\partial}^2}{2 m_t}+i\Gamma_t
\right)\chi
\nonumber\\
&&\hspace*{-2.0cm}
+\, \int d^{d-1} r \left[\psi^\dagger\psi\right](r)
\left(-\frac{C_F\alpha_s}{r}\right) 
\left[\chi^\dagger\chi\right](0)
\nonumber\\
&&\hspace*{-2.0cm}
+ \, \mbox{NLO}.
\end{eqnarray}
The leading order Coulomb potential is part of the leading order 
Lagrangian and cannot be treated as a perturbative interaction term 
as anticipated above. At NNLO the two-loop correction to the 
Coulomb potential \cite{Sch98} is required, as well as  
the potentials 
of the form $\alpha_s^2/r^2$, $\alpha_s/r^3$ and $\alpha_s\delta(r)$ 
and kinetic energy corrections. A remarkable feature is that ultrasoft 
gluon interactions contribute only from NNNLO on. There are 
no gluon (light quark) fields in the PNRQCD Lagrangian to  
NNLO. It is after matching NRQCD on PNRQCD that the computation of the 
threshold cross section turns into an essentially quantum-mechanical 
problem. Nevertheless, matching is crucial. The 
non-Coulomb potentials result in ultraviolet divergent integrals which 
we regulate dimensionally (the potentials are therefore needed in 
$d$ dimensions -- see \cite{BSS99}). The resulting regulator-dependence 
cancels with the regulator-dependence of the matching coefficients 
in NRQCD and PNRQCD. (This provides a check of the logarithm in the 
2-loop contribution to $c_1$ and verifies the 2-loop anomalous dimension 
of the non-relativistic current computed in \cite{BSS98} from the 
effective theory side.)

The potentials should be considered as short-distance coefficients 
of non-local, instantaneous operators in the PNRQCD Lagrangian. This 
implies that the potentials are infrared finite, but can contain 
(parametrically large) logarithms $\ln(\mu_s/\mu_{us})$, where 
$\mu_s\sim m_t v$ is the scale at which NRQCD is matched on PNRQCD and 
$\mu_{us}\sim m_t v^2$ is the renormalization scale of PNRQCD. Explicit 
calculation of the potentials to NNLO demonstrates that they do not 
contribute LL terms. An example of a NNLL term is provided by the 
logarithm in the Coulomb potential at order $\alpha_s^4$ 
\cite{ADM78}.

\subsection{Step 3: PNRQCD perturbation theory}

Perturbation theory in PNRQCD has much in common with perturbation 
theory in quantum mechanics. The lowest order Coulomb potential is 
part of the unperturbed Lagrangian. Perturbation theory in PNRQCD 
does not use ordinary free (non-relativistic) quark propagators, 
but a propagator for a $t\bar{t}$ pair in the presence of the 
Coulomb interaction. The corresponding Coulomb Green function 
$G_c(\bff{r},\bff{r}^\prime;\bar{E})$ solves 
\begin{eqnarray}
\label{schroedinger}
&&\hspace*{-0.65cm}
\left[-\frac{\bff{\nabla}^2_{\!\bff{r}}}{m_t} + V(\bff{r}) 
- \bar{E}\right] G_c(\bff{r},\bff{r}^\prime;\bar{E}) = 
\delta^{(3)}(\bff{r}-\bff{r}^\prime), 
\nonumber\\[-0.2cm]
&&
\end{eqnarray}
where $\bar{E}=E+i\Gamma_t$ and $V(\bff{r})=-C_F\alpha_s/r$. (More 
precisely, the dimensionally regularized Coulomb Green function is 
defined through the corresponding integral equation in momentum space.) 
The non-leading potentials can be treated perturbatively, which leads 
to integrals of the form
\begin{eqnarray}
&&\hspace*{-0.65cm}
\int\! \frac{d^3\bff{p}}{(2\pi)^3} \frac{d^3\bff{p}^\prime}{(2\pi)^3} 
\frac{d^3\bff{q}_1}{(2\pi)^3}\frac{d^3\bff{q}_2}{(2\pi)^3} \,\,
\tilde{G}_c(\bff{p},\bff{q}_1;\bar{E})\nonumber\\
&&\hspace*{0.0cm}
\cdot\,\delta V(\bff{q}_1-\bff{q}_2)
\cdot \,\tilde{G}_c(\bff{q}_2,\bff{p}^\prime;\bar{E})
\end{eqnarray}
and generalizations with more than one insertion of an interaction 
potential $\delta V$. 
Alternatively, one can solve the Schr\"odinger equation 
(\ref{schroedinger}) with the NNLO potential rather than the LO 
potential exactly. The two methods are equivalent at NNLO, but differ 
by higher order terms. This completes the calculation of the resummed 
cross section.

\section{TOP QUARK MASSES}

A plot of the threshold cross section displays a large NNLO 
correction -- see the lower panel of Fig.~\ref{fig1} below -- that 
affects both the normalization of the cross section and the 
position of what is left over from the 1S resonance. It therefore 
impacts directly on the accuracy with which the top quark mass 
can be determined from a future measurement. It is worth thinking 
about the origin of this large correction \cite{Ben98}.

We have implicitly assumed that the top quark mass is renormalized 
on-shell. The NRQCD Lagrangian (\ref{nrqcd}) refers to this choice, 
but we could have added a small mass term $\delta m_t \psi^\dagger\psi$. 
The only requirement is that $\delta m_t \sim m_t v^2$ or smaller, 
so that it is of the same order or smaller than the leading order terms 
in the NRQCD Lagrangian. This option turns out to be useful in combination 
with the observation that the large NNLO correction to the cross 
section peak position is caused by large perturbative 
corrections to the coordinate space Coulomb potential in the 
Schr\"odinger equation. The large corrections arise from loop momentum 
smaller than $m_t v$ and we can absorb them into the 
quantity
\begin{equation}
\label{masssub}
\delta m_t(\mu_f) = -\frac{1}{2}\int\limits_{|\vec{q}\,|<\mu_f} 
\!\!\!\frac{d^3\bff{q}}{(2\pi)^3}\,[\tilde{V}(q)]_{\rm Coulomb},
\end{equation}
where $[\tilde{V}(q)]_{\rm Coulomb}$ is the Coulomb potential in 
momentum space. Then define the subtracted potential 
\begin{equation}
V(r,\mu_f)=V(r)+2\delta m_t(\mu_f),
\end{equation}
which should have smaller perturbative corrections.
Since $\delta m_t(\mu_f)$ is $r$-independent, it is a legitimate mass 
subtraction, provided we choose the subtraction scale at most of 
order $m_t v$. (Beyond NNLO the Coulomb potential contains a logarithm 
as discussed above. To define $\delta m_t(\mu_f)$ completely beyond 
NNLO, one then has to specify the scale of this logarithm in addition to 
$\mu_f$.) Hence we rewrite the Schr\"odinger equation (\ref{schroedinger}) 
identically in terms of the subtracted potential and $E=\sqrt{s}-2\,
m_{t,\rm PS}(\mu_f)$, where 
\begin{equation}
\label{psmass}
m_{t,\rm PS}(\mu_f) \equiv m_t-\delta m_t(\mu_f)
\end{equation}
defines a new renormalized mass parameter, the potential-subtracted 
(PS) mass \cite{Ben98}. 

This reparametrization of the threshold cross section in terms of the 
PS mass rather than the pole mass should remove the large higher-order 
corrections to the peak position -- and we shall see later that it does. 
The real benefit, however, 
is that the improved convergence should occur as well 
in other processes involving nearly on-shell top quarks. The argument 
goes as follows: the integrals that relate the top quark pole mass 
to the $\overline{\rm MS}$ mass (or bare mass, for that matter) also 
lead to large higher-order corrections from loop momentum smaller 
than $m_t$ \cite{BB94}. One can show by explicit calculation at 
one loop \cite{Ben98,HSSW98}, and by diagrammatic arguments in higher 
orders \cite{Ben98}, that the dominant 
corrections to $m_t$ and the coordinate space Coulomb potential are 
identical, so that they cancel between $m_t$ and $\delta m_t(\mu_f)$ 
in (\ref{psmass}), when $m_{t,\rm PS}(\mu_f)$ is related to the 
$\overline{\rm MS}$ mass 
$\bar{m}_t=m_{t,\overline{\rm MS}}(m_{t,\overline{\rm MS}})$. Hence 
any process that is not as sensitive to long-distance corrections as 
the pole mass and the coordinate space potential separately is 
expected to have smaller perturbative corrections when expressed 
in terms of the PS mass instead of the pole mass. (Of course, in 
processes which involve top quarks only far off mass-shell one can use 
$\bar{m}_t$ directly.)

Explicitly, we find the following numerical 
expressions for the series that relate the pole and PS mass, respectively, 
to the $\overline{\rm MS}$ mass $\bar{m}_t$ ($\bar{m}_t=165\,$GeV and 
$\alpha_s(\bar{m}_t)=0.1083$):
\begin{eqnarray}
&&\hspace*{-0.65cm}
m_t = \big[165.0+7.58+1.62+0.51
\nonumber\\
&&+\,0.24\,(\mbox{est.})\big]\,
\mbox{GeV}
\label{polerel}\\
&&\hspace*{-0.65cm}
m_{t,\rm PS}(20\,\mbox{GeV}) = \big[165.0+6.66+1.20+
0.28\nonumber\\
&&+\,0.08\,(\mbox{est.})\big]\,
\mbox{GeV}.
\label{psrel}
\end{eqnarray}
The 3-loop coefficients are computed using \cite{CS99}. The 4-loop estimate 
uses the `large-$\beta_0$' limit \cite{BBB}. (Note that a NNLO calculation 
of the threshold cross section determines the PS mass with a parametric 
accuracy $m_t \alpha_s^4$. To profit from this accuracy for the 
$\overline{\rm MS}$ mass requires the 4-loop perturbative relation.) 
The improved convergence is evident and significant on the scale 
of $0.1\,$GeV set by the projected statistical uncertainty on the 
mass measurement.

The arguments above show that it is advantageous to abandon the 
on-shell mass renormalization scheme even for pair production 
near threshold. This is perhaps one of the most important conclusions 
that emerged from the NNLO calculations. In the following we will use 
the PS scheme with $\mu_f=20\,$GeV. Other choices of $\mu_f$ are 
conceivable. Using the definitions (\ref{masssub},\ref{psmass}), the PS 
masses at different $\mu_f$ are easily related.

The PS scheme could be called a `minimal' subtraction scheme, because it 
subtracts the large infrared correction, but not more. Other,  
`non-minimal', mass definitions can be conceived that fulfil the same 
purpose (see, for instance, the second reference of \cite{nnlo2}).

\section{RESULT}

\begin{figure}[t]
   \vspace{-2.3cm}
   \epsfysize=9.5cm
   \epsfxsize=7cm
   \centerline{\epsffile{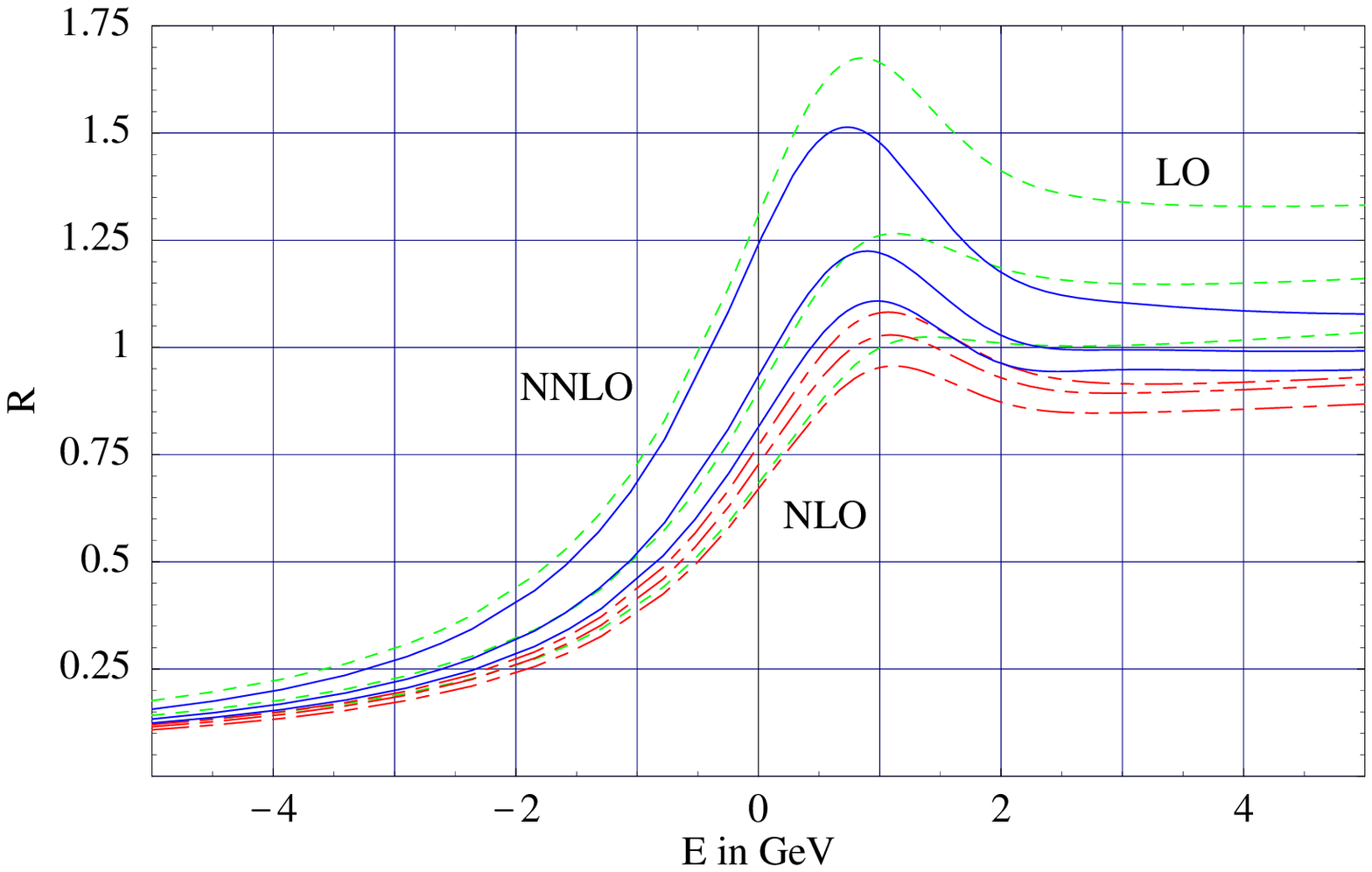}}
   \vspace*{-5.0cm}
   \hspace*{-0.1cm}
   \epsfysize=9.5cm
   \epsfxsize=7cm
   \centerline{\epsffile{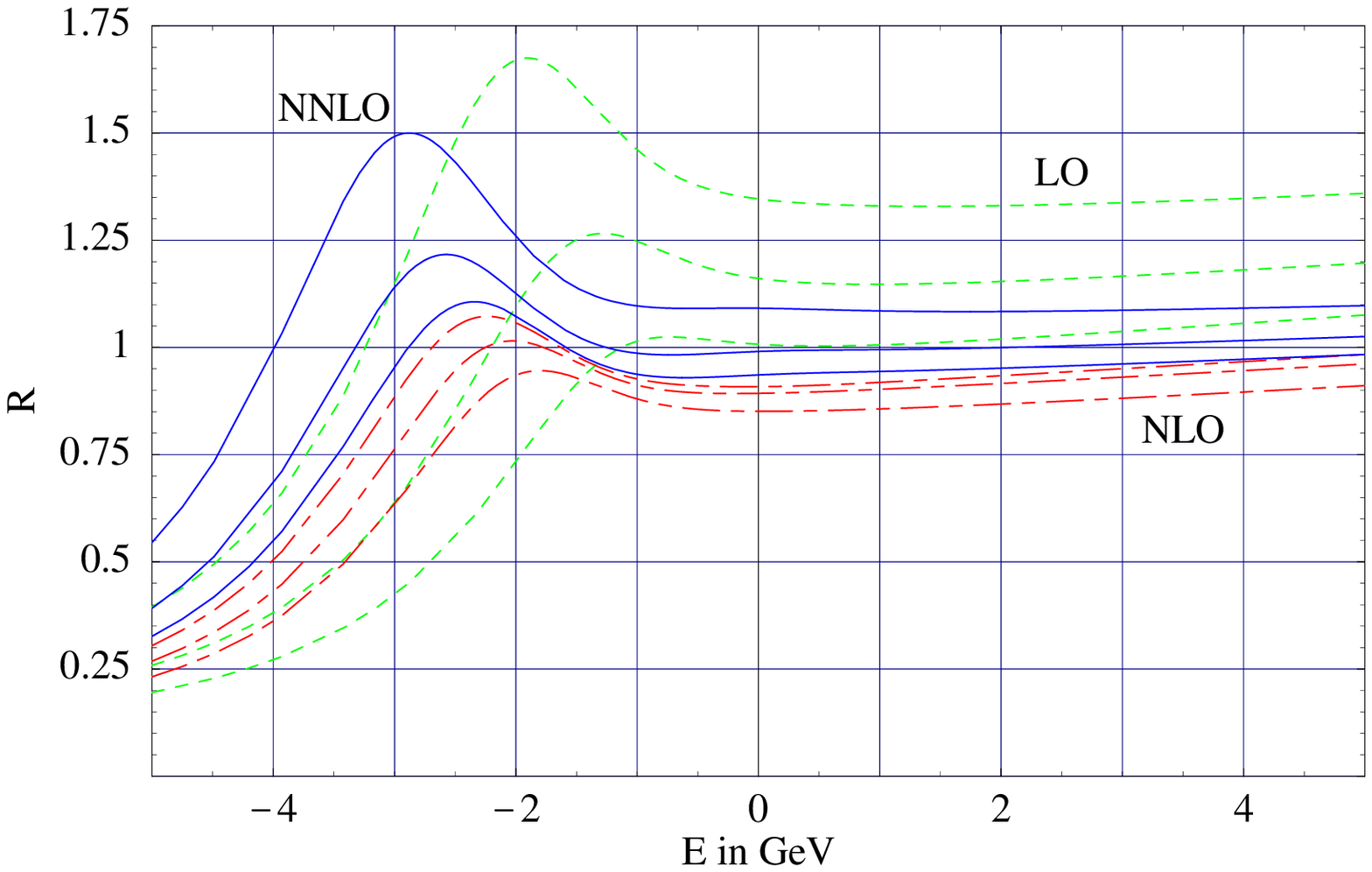}}
   \vspace*{-3.8cm}
\caption[dummy]{\label{fig1}\small 
(a) [upper panel]: The normalized $t\bar{t}$ cross section (virtual 
photon contribution only) in LO (short-dashed), NLO (short-long-dashed) 
and NNLO (solid) as function of 
$E=\sqrt{s}-2 m_{t,\rm PS}(20\,\mbox{GeV})$ (PS scheme, $\mu_f=20\,$GeV).
Input parameters: $m_{t,\rm PS}(20\,\mbox{GeV})=\mu_h=175\,$GeV, 
$\Gamma_t=1.40\,$GeV, 
$\alpha_s(m_Z)=0.118$. The three curves for each case refer to 
$\mu=\left\{15 (\mbox{upper}); 30 (\mbox{central}); 
60 (\mbox{lower})\right\}\,$GeV. (b) [lower panel]: 
As in (a), but in the pole mass scheme. Hence 
$E=\sqrt{s}-2 m_t$. Other parameters as 
above with $m_{t,\rm PS}(20\,\mbox{GeV})\to m_t$.}
\end{figure}

The top quark cross section at LO, NLO and NNLO (including the 
summation of logarithms at NLL) is shown in the upper panel 
of Fig.~\ref{fig1}. (To be 
precise, the NLO curves include the second iterations of the 
NLO potentials.) For comparison, the result in the on-shell scheme 
is shown in the lower panel.

In the PS scheme, contrary to the on-shell scheme, the peak 
position varies little, if consecutive orders in the expansion 
are added. 
On the other hand, the NNLO correction to the peak height 
is large, and a significant 
uncertainty of about $\pm 20\%$ in the normalization remains,  
larger than at NLO. 

The strong enhancement of the peak for the small 
scale $\mu=15\,$GeV is a consequence of the fact that the perturbative 
corrections to the residue of the 1S pole  
become uncontrollable at scales not much smaller than $15\,$GeV. 
These large corrections are mainly associated with the 
logarithms that make the coupling run in the Coulomb potential. This 
could be interpreted either as an indication that higher order 
corrections are still important (at such low scales) or that the terms 
associated with $b_0$ should be treated exactly, because they are 
numerically (but not parametrically) large.

If we take (naively) the change in the peak position under scale 
variations as a measure of the uncertainty of the top mass measurement, 
we conclude that a determination of the PS mass with an error of 
about $100\,$-$150\,$MeV is possible. Given that the uncertainty in 
relating the PS mass to the $\overline{\rm MS}$ mass is of the 
same order -- see (\ref{psrel}) --  this accuracy seems to be sufficient. 
It is worth emphasizing that this conclusion does not hold for the 
top quark pole mass, see (\ref{polerel}) and the lower panel of 
Fig.~\ref{fig1}. For a realistic assessment of the error in the 
mass measurement at a future high-energy lepton collider, the theoretical 
line shape has still to be folded with initial state radiation, beamstrahlung 
and beam energy spread effects. Since these effects are well understood, 
the main question that needs to be addressed is whether the normalization 
uncertainty leads to a degradation of the mass measurement after 
these sources of smearing are taken into account. This should be studied 
in a collider design specific setting.

\section{BEYOND NNLO AND OPEN QUESTIONS}

The recent developments have put the calculation of the threshold cross 
section on a more systematic basis. While increasing the parametric 
accuracy to NNLO, and addressing the issue of mass renormalization in 
this context for the first time, they have also shown that the 
theoretical uncertainties are larger than what has commonly been 
assumed.

The normalization uncertainty is particularly disconcerting. It suggests 
that yet higher orders in the resummed expansion could be 
important. One should try to understand whether these large 
corrections have a physics origin, whether they can be resummed or whether 
they can be eliminated similar to the large corrections to the 
peak position in the on-shell scheme.

A complete NNNLO calculation appears to be prohibitive by present 
standards, primarily because it requires the 3-loop coefficient function 
of $\psi^\dagger\sigma^i \chi$ and the 3-loop Coulomb potential. 
But it is already interesting (and possible) to 
address a well-defined subset of terms. The most obvious subset concerns  
ultrasoft (retardation) effects, which occur for the first time 
at NNNLO. They are interesting, because they introduce an explicit 
sensitivity to the scale $m_t v^2\sim 2\,$GeV and the strong coupling 
normalized at this small scale. 

The sensitivity to the ultrasoft 
scale exists already in the NLL approximation, in which logarithms 
sensitive to the ultrasoft scale are summed, with no ultrasoft diagrams 
to be computed. In addition to the missing inputs to the NRQCD 
renormalization group, this requires understanding the renormalization 
group scaling of PNRQCD, which has not been addressed so far. 
$\alpha_s(m_t v^2)$ then appears as the endpoint of the 
renormalization group evolution.

Diagrams with one ultrasoft gluon in the 
Coulomb background represent a true NNNLO correction of order 
$\alpha_s(m_t v^2) v^2$. The corresponding correction to the 
1S toponium energy level and wave function at the origin has 
already been computed \cite{KP99}. Near the peak position this 
constitutes the dominant contribution to the cross section. 
To assess the numerical significance of this correction, it is necessary 
to combine it with the NNLL terms that cancel the regulator-dependence 
of the ultrasoft diagrams.

Accounting for the top quark width correctly represents another 
challenge. It is an interesting theoretical problem by itself to generalize 
a non-relativistic effective theory description to the threshold 
production of unstable particles. Its solution may be useful elsewhere. 
For the particular case of top quarks, this leads us outside the 
well-defined framework where only QCD corrections may be discussed. 
Beyond the leading order implementation of the width currently 
adopted, one has to consider a more general set of one-loop electroweak 
corrections. In addition, non-factorizable corrections due to 
decay products of the top quark interacting with the other top quark or 
its decay products, and diagrams with single-resonant top quarks, 
cannot be neglected; the problem has to be formulated in terms 
of a particular final state such as $WWb\bar{b}$. However, we also 
expect that these corrections are `structure-less', that is, do not 
exhibit a pronounced resonance peak. For this reason, we anticipate 
that they add to the already existing normalization uncertainty, but 
affect little the top quark mass measurement.

\section*{ACKNOWLEDGEMENTS}

This summary is based on work done in collaboration with A.~Signer 
and V.A.~Smirnov. It is supported in part by the 
EU Fourth Framework Programme `Training and Mobility of
Researchers', Network `Quantum Chromodynamics and the Deep Structure 
of Elementary Particles', contract FMRX-CT98-0194 (DG 12 - MIHT).

\end{document}